%% file: main.tex
\documentclass[fleqn,usenatbib]{mnras}

\usepackage[T1]{fontenc}
\usepackage{ae,aecompl}

\usepackage{graphicx}	
\usepackage{amsmath}	
\usepackage{amssymb}	

\usepackage[figure,figure*]{hypcap}
\usepackage{rotating}

\input{personal_set}
\newcommand{\stara}{KIC\,3858884}
\newcommand{\starb}{KIC\,4544587}
\newcommand{\starc}{KIC\,10661783}
\newcommand{\stard}{HD\,172189}
\newcommand{\stare}{CID\,100866999}
\newcommand{\starf}{CID\,105906206}
\newcommand{\starg}{HD\,159561}
\newcommand{\starh}{KIC\,9851944}
\newcommand{\stari}{KIC\,8262223}
\newcommand{\starj}{KIC\,10080943}
\newcommand{\stark}{HD\,15082}
\newcommand{\starl}{HD\,174966}
\newcommand{\starm}{HD\,174936}
\newcommand{\forb}{$\nu_{\mathrm{orb}}$}

\usepackage{newtxtext,newtxmath}

\title[Rotational Splittings in \dss]{Study of rotational splittings in $\delta$~Scuti stars using pattern finding techniques}

\author[Ram\'on-Ballesta et al.]{A. Ram\'on-Ballesta$^{1,2,3}$\thanks{\href{mailto:a.ramonballesta@gmail.com}{a.ramonballesta@gmail.com}, \href{mailto:arb@iaa.es}{arb@iaa.es}},
A. \gh$^{2,1}$,
J. C. Su\'arez$^{2,1}$,
J. R. Rod\'on$^{1}$,
\newauthor J. Pascual-Granado$^{1}$,
R. Garrido$^{1}$
\\
$^{1}$Instituto de Astrof\'{\i}sica de Andaluc\'{\i}a (CSIC), Glorieta de la Astronom\'{\i}a S/N, 18008, Granada, Spain\\
$^{2}$Universidad de Granada, Dept. Theoretical Physics and Cosmology, 18071, Granada, Spain\\
$^{3}$Universidad Internacional de Valencia, E-46021, Valencia, Spain\\
}

\date{Accepted XXX. Received YYY; in original form ZZZ}

\pubyear{2021}

\begin{document}
\label{firstpage}
\pagerange{\pageref{firstpage}--\pageref{lastpage}}
\maketitle

\begin{abstract}
Detecting and understanding rotation in stellar interiors is nowadays one of the unsolved problems in stellar physics. Asteroseismology has been able to provide insights on rotation for the Sun, solar-like stars, and compact objects like white dwarfs. However, this is still very difficult for intermediate-mass stars. These stars are moderate-to-rapid rotators. Rotation splits and shifts the oscillation modes, which makes the oscillation spectrum more complex and harder to interpret. Here we study the oscillation patterns of a sample of benchmark \dss\ belonging to eclipsing binary systems with the objective to find the frequency spacing related to the rotational splitting (\dr). For this task, we combine three techniques: the Fourier transform, the autocorrelation function, and the histogram of frequency differences. The last two showed a similar behaviour. For most of the stars, it was necessary to determine the large separation (\Dnu) prior to spot \dr. This is the first time we may clearly state that one of the periodicities present in the p~modes oscillation spectra of \dss\ corresponds to the rotational splitting. This is true independently of the stellar rotation rate. These promising results pave the way to find a robust methodology to determine rotational splittings from the oscillation spectra of \dss\ and, thus, understanding the rotational profile of intermediate-mass pulsating stars.
\end{abstract}




\section{Introduction}
\label{sec:intro}

The \dss\ are intermediate-mass ($\sim$1.5 and 3 \msol) pulsators located in the instability strip of classical pulsators (e.g. Cepheids) but with spectral types ranging from A to F. They can be found on the main sequence or burning hydrogen in the shell. They are moderate-to-fast rotators \citep[see][and references therein]{royer2007}. Measurements of surface projected velocities, \vsini, and statistical inferences on actual rotation rates show that they can reach values close to the break-up limit. Thus, these stars are very good laboratories to test theories of angular momentum and chemical transport in stellar interiors. However, rotation severely hampers an accurate determination of stellar global parameters, such as effective temperature and surface gravities. To solve this problem it is necessary to determine the actual rotational velocity of these stars.\par

Up to date, only interferometric techniques for bright and deformed stars \citep{vanBelle2001} may be able to determine the angle of inclination of the star, and therefore, the actual rotation velocity. High-resolution spectroscopy may provide us with an estimate of the inclination of fast rotators with low-to-moderate \vsini\ by modelling the gravitational darkening \citep{Fremat2005} or analysing line-profile variations \citep{aerts1993}.\par

\begin{table*}
    \centering
    \input{table_01.tex}
\end{table*}

Thanks to space missions like \most\ \citep{walker2003}, \corot\ \citep{baglin2006}, \kepler\ \citep{koch2010}, and \tess\ \citep{Ricker2015}, significant progress has been made in the determination of the internal rotation profiles through the study of g modes \citep{vanreeth2015,vanreeth2016}. Unfortunately, this has not been achieved for surface rotation potentially provided by p-modes of intermediate-mass stars. Even so, it has been possible to detect periodic patterns in the p-mode frequency spectra of \dss\ (\citealp[see e.g.][]{GH09, GH13,mantegazza12}; \citealp[or, more recently,][]{bedding2020,murphy2021}). These patterns were predicted to be a large separation, \Dnu, in the low radial order regime \citep{suarez2014}, in contrast to the well-known large separation found in the asymptotic regime of solar-like stars. It was then empirically corroborated by the scaling relation found between the stellar mean density of a sample of benchmark eclipsing binaries and the observed large separation \citep{GH15, GH17}. Complementary theoretical works supported the discovery that such scaling relation was almost independent of the rotation rate \citep{reese2008, ouazzani2012, Mirouh19}. Therefore, today the low-order \Dnu\ is key to any asteroseismic study of \dss.\par

Although the aforementioned space missions have provided us with ultra-high precision photometric data, most of their targets are too faint to have their inclination determined by applying the spectral or interferometric analysis. In this work, we focus on the study of the rotation frequency of \dss\ from its fingerprints on their oscillation spectrum. The frequency pattern is affected by the rotation, breaking the degeneracy in the azimuthal order and introducing an additional pattern in the frequency spectrum \citep[see, e.g.,][]{goupil2005} called rotational splitting (hereon \dr). The visibility of some modes is also affected by the angle of inclination \citep{suarez2002,casas2006} and, in fast rotators, the effect of the rotation can even redistribute the modes \citep{lignieres2006}, although both \Dnu\ and \dr\ might be still detectable \citep{reese2017}. It is also possible for \dr\ to combine with \Dnu\ in the frequency spacings \citep{paparo2016-I,paparo2016-II,barcelof2017,guo2017}. Thus, finding \dr\ would make it a valuable observable to determine the rotation of \dss\ in an independent way. We aim at developing a methodology to detect \dr\ of \dss\ just from their oscillation spectra.\par

The paper is organised as follows. We describe the observations in Sect.~\ref{sec:sample}, and the methodology in Sect.~\ref{sec:methodology}. Discussion of the results is given in Sect.~\ref{sec:results}. We applied what we have learned from the results to stars for which rotation is unknown in Sect.~\ref{sec:unknownrotation}. Conclusions and future perspectives are explained in Sect.~\ref{sec:conc}. Finally, for the sake of completeness, we provide the figures for \starb\ and \starh\ in the appendix \ref{sec:appendix} since we are not citing them in the main text.\par

\section{The observational sample} \label{sec:sample}

In order to develop a methodology to find patterns in the oscillation spectra linked to \dr, we need a well characterised sample for which the true rotational velocity has been determined. This is a really hard task because of the difficulties in measuring stellar inclination angles. An approximate estimate can be derived for binary systems. That is why we used the benchmark \dss\ in eclipsing binary systems published by \citet{GH17}. Assuming that the rotation and system axes are parallel, we can get the surface rotation of a star. This assumption is widely used in the literature and is likely to be true in the majority of the cases of eclipsing binaries, as shown in Table 1 of \citet{albrecht2011}. However, some exceptions are known to happen in some binary systems \citep[see, e.g.][]{albrecht2009, albrecht2014} and, especially, in several hot-Jupiters hosts \citep[see, e.g.][]{hebrard2008, winn2009, narita2009, triaud2010}. These examples must be taken into account as a possible explanation for the cases in which we obtain unexpected results, especially \stark\ (see Sect~\ref{subsec:peculiar}).\par

\begin{figure}
	\includegraphics[width=\columnwidth]{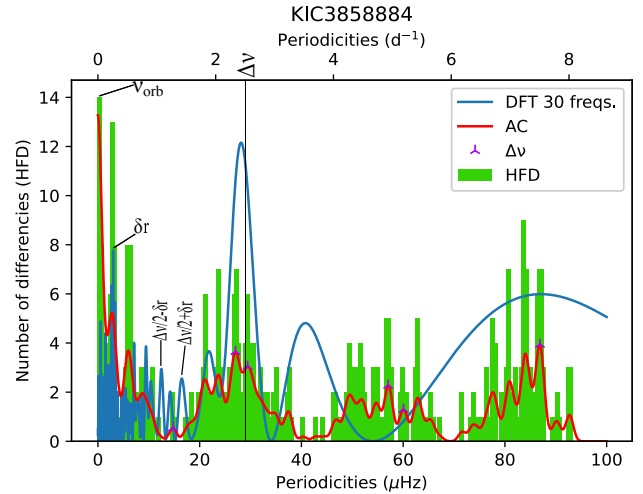}
    \caption{\stara. The peaks for \Dnu\ and \dr\ can be easily identified in the DFT (blue line), as well as the combinations \Dnu/2$\pm$\dr. The HFD (green bars) shows the expected grouping around \Dnu, 2\Dnu\ and 3\Dnu\ as well as the peaks corresponding to \forb\ and \dr. The AC function (red line) shows the peaks corresponding to \Dnu, 2\Dnu\ and 3\Dnu\ (purple cross).}
    \label{fig:stara}
\end{figure}

\begin{figure}
	\includegraphics[width=\columnwidth]{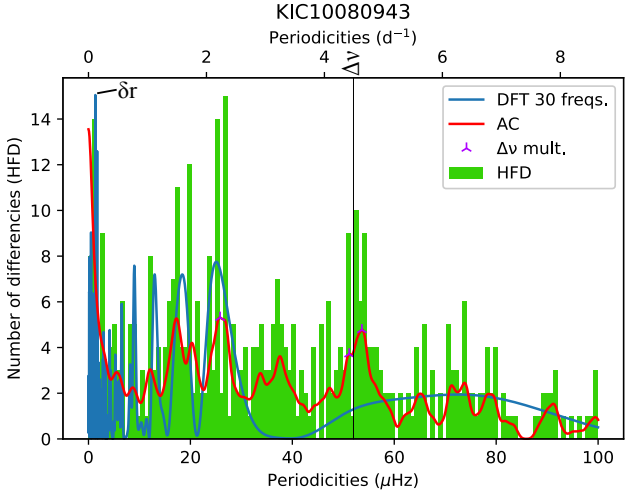}
    \caption{\starj. In the DFT of the main frequencies the peaks corresponding to \Dnu/2, \Dnu/3 and, especially, \dr\ can be seen. The HFD shows \dr\ and a grouping around \Dnu. Also, the peak corresponding to \Dnu/2 is highly noticeable above all. The AC function shows a peak for \Dnu.}
    \label{fig:starj}
\end{figure}

This sample contains 9 eclipsing binary systems, a \ds\ star harbouring a planet and a triple system for which the radius of the main pulsating component has been accurately determined with optical interferometry. Precise stellar parameters and system inclination angles are provided in the literature (see Table~\ref{tab:sample}). With all this information, we estimated the expected rotational velocities and hence the corresponding rotational splittings. Note that the radii shown in Table~\ref{tab:sample} correspond to the mean radius as explained in \citet{GH15,GH17}\footnote{This radius is calculated using the real volume of the star, i.e., corrected for the effect of rotational deformation taking a Roche model, but for a spherical model. This is then the radius of a sphere with the same volume as the real spheroidal star.}. Therefore, given that the equatorial radius is bigger than the mean radius, the calculated \dr's would be a slight overestimate of the true splittings. Although the sample covers a wide range of rotations, from ${\sim}25$~\kms\ (\stara) to ${\sim}239$~\kms\ (\starg), the actual surface velocities remain in an average of ${\sim}78$~\kms\ providing \dr$\sim$[2, 19] \muhz.\par

In addition, \citet{GH17} identified the low-order large separations (\Dnu) for the \dss\ of these systems. This quantity is crucial to disentangle \Dnu\ and \dr\ patterns (see next Sect.). This paper is focused on finding \dr, hence we have directly taken the values of \Dnu\ from \citet{GH17}.\par

\section{Methodology} \label{sec:methodology}

The first step is to only select modes in the regime where the surface \dr\ appears, i.e., pressure modes, as well as to clean the oscillation spectra from combinations and harmonics (in particular those coming from binarity). Additionally, other artefacts like tidal forces, mass transfer or the effects of the eclipses may remain. We assume these later as negligible, although they might become important when determining \dr, especially when their temporal scale is similar to the stellar rotation period.\par

When several periodic patterns are expected to be found, it is important to understand their nature and domain. Low-order large separations in \dss\ may vary from $80-100$ \muhz\ at the ZAMS to $10-20$ \muhz\ for stars at the end of the main sequence or during the post-main sequence \citep{rodriguezj2020,suarez2014}. This is the range we focus here although, according to \citet{evano2019a,evano2019b}, chaotic modes may have non-negligible mode amplitudes showing pseudo-regularities involving a frequency separation close to \Dnu. This would, indeed, strengthen \Dnu\ signature. Note that for evolved fast rotators \Dnu\ is of the order of the rotation frequency. In the hypothesis that both \Dnu\ and \dr\ are present in the oscillation spectrum as independent periodicities, such an overlap may hamper their detection. Other effect that may also hide the detection of \dr\ is the well-known rotation-pulsation interaction at high rotation rates, which modifies the rotational splittings by making them uneven \citep[see e.g.][]{soufi1998,Suarez2006,reese2009} or, to a greater extent, causes a rearrangement of the pulsation spectra \citep[][]{lignieres2006}. This, combined with the visibility of the modes (see Sect.~\ref{sec:intro}), may result in frequency patterns that mimic the \Dnu\ signature \citep[see e.g.][]{GH13} or become a linear combination of \Dnu\ and \dr\ \citep{paparo2016-I,paparo2016-II}.\par

\begin{figure}
	\includegraphics[width=\columnwidth]{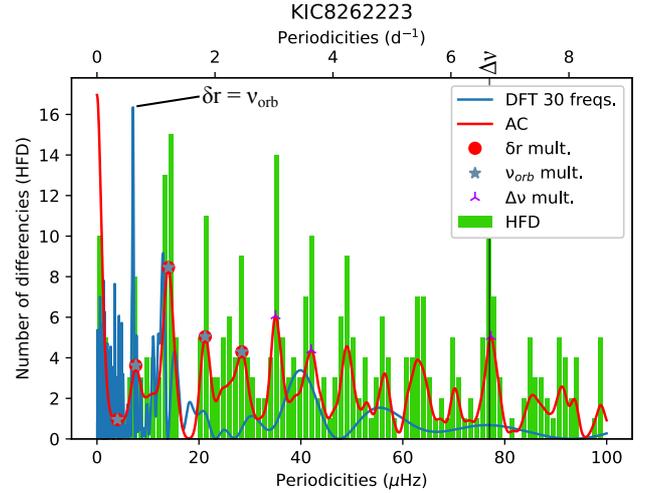}
    \caption{\stari. The orbital synchronisation and the coincidence that \Dnu$\simeq$11\dr\ (within the error bars) could be the reason for so many multiples of \dr\ in the HFD.}
    \label{fig:stari}
\end{figure}

Up to date, we have no clue about which of the traditional methods to search for patterns is better suited to study the different cases mentioned above. We chose the most efficient ones to determine \Dnu\ and applied them simultaneously. These are:

\begin{itemize}
    \item [$\bullet$] The discrete Fourier transform (DFT, from now on) of the normalised frequency spectrum, as proposed by \citet{GH09} and subsequent papers. This method assumes that the frequencies with the highest amplitudes are those which primarily contribute to the large separation, because they would mainly correspond to low degree modes. Then, the frequencies amplitudes are made equal to one before computing the DFT. The DFT method, thus, restricts its application to the highest amplitude frequencies (generally 30-40). Therefore, such a restriction may hamper the search for the rotational splitting simultaneously. On the other way around, if DFT is applied to all the observed frequencies, those which do not contribute to \Dnu\ may blur the DFT results. Note that we are drawing the DFT using 1/\textit{t} in the abscissa instead of the usual way of using \textit{t}. We think that this may be beneficial to show the low periodicities that would correspond to \dr.\par

    \item [$\bullet$] The autocorrelation function (AC, from now on). This well-know technique is generally useful to spot patterns that consist in a succession of (shifted) copies of themselves. We find this technique to be quite efficient in finding \Dnu\ and \dr\ (see Fig. \ref{fig:stara}). This method has already been used to extract \Dnu\ in rapidly rotating models by \citet{reese2017}, who found that in combination with DFT it is possible to distinguish between \Dnu\ and \dr\ in some specific stellar configurations and rotation rates.\par

    \item [$\bullet$] The histogram of frequency differences (HFD, from now on). This method was previously used to find \Dnu\ by \citet{handler1997}. The frequency pattern due to rotational splitting is expected to be replicated (to some extent) around some frequencies because $m=[-\ell, \ell]$. This makes the HFD suitable to complement DFT because the difference corresponding to \dr\ should appear as a prominent peak in such analysis.\par
\end{itemize}

We have performed the following procedure. First, we have computed the DFT, AC, and HFD for all the stars of the sample. Once \Dnu\ signature had been identified, in each of the 3 transforms we searched for a peak corresponding to the expected value of \dr\ that we estimated using the physical parameters derived from the orbital solution. We investigated other possibilities, such as this peak could be the double of \dr\ \citep{reese2017} or a linear combination of \dr\ and \Dnu\ \citep{paparo2016-II}.\par

Finally, we have arbitrarily scaled the DFT an AC curves in order to fit them within the HFD figure in a reasonable size.\par

\section{Results}
\label{sec:results}

We identified several groups depending on the behaviour of their HFD, DFT and AC. They cannot form any real category, since we have a small sample, but could be indicative of them.\par

\begin{itemize}
    \item[$\bullet$] \textbf{Favourable}. This group is composed of stars for which \Dnu\ and \dr\ can be easily identified in at least two of the diagnostics diagrams. For this small sample, this occurs for DFT and HFD. The AC is not able to detect the splitting due to its limited resolution at periodicities near 0. This group gathers two stars with the lowest rotational splittings in the sample.
    
     \item[$\bullet$] \textbf{Synchronised}. This second group assembles the stars whose rotation is synchronised with the orbital period, which makes their disentanglement difficult.
     
     \item[$\bullet$] \textbf{Evolved}. In a third group we have gathered together the stars with high $(R/\mathrm{R}_{\odot})/(M/\mathrm{M}_{\odot})$ ratio, which are likely post-main sequence stars and, therefore, the most evolved ones of the sample. At this evolutionary stage, mixed modes come up, blurring any periodicity present in the oscillation spectrum.
     
     \item[$\bullet$] \textbf{Peculiar}. The remaining stars correspond to peculiar cases. They either do not fit in any of the previous groups or have particular characteristics.
    
\end{itemize}

\subsection{Favourable cases}
\label{subsec:favourable}
For \stara\ and \starj, we clearly identified the theoretical \dr\ of 1.9 and 1.7 \muhz\, respectively. Both show a peak in the HFD (Figures~\ref{fig:stara} and \ref{fig:starj}), independently of \Dnu\ and \forb. The peak is seemingly confirmed by their DFT diagnostic as well. For this latter, it is worth noticing that the \dr\ peak for \stara\ is especially significant when using the first 60 highest-amplitude frequencies. Interestingly, for determining \Dnu\ with this method, it has been shown \citep[e.g.][]{GH09,GH13} that 30 frequencies are generally sufficient to show the relevant peak with a significant power. In that case the radial and a few $\ell=1$ mode frequencies are assumed to be contributors to the \Dnu\ peak. However, it is expected that the main contributors for \dr\ are rather non-radial modes and their $m$ components, and hence, the overall number of frequencies is logically increased.\par

On the other hand, the autocorrelation function confirms the detection of \Dnu\ and \dr. The autocorrelation function works this well in the majority of the cases for the detection of \Dnu\ but this is not always true for \dr.\par

Both stars are those with the highest number of frequencies and the slowest rotation of the entire sample. The first fact leads us to think that the highest number of frequencies statistically improves the chances of having the best ones for identifying \Dnu\ and \dr\, although this does not work so well in some cases (e.g. \starm\ and \starl\ in Sec. \ref{sec:unknownrotation}). The slow rotation, making \Dnu$\gg$\dr, might favour the detection as well (even though there was a weak chance of also detecting a combination of \Dnu\ and \dr). Another advantage here is the inclination angle of \stara, nearly equator-on, which increases the amplitude of the even modes, especially \ele=2 \citep[see Fig. 11 from][]{reese2017}. This might explain the comb form of the HFD in this particular case with multiples of \dr\ (see Fig.~\ref{fig:stara}).\par

\subsection{Orbital synchronisation}
\label{subsec:synchro}
The stars with orbital synchronisation are \starb, \starh, and \stari. In the last two, the rotation is synchronised with the orbital period in 1:1 resonance. This makes the expected \dr\ indistinguishable from the orbital frequency \forb. In \starb, \dr\ is roughly the double of \forb.\par

In all three cases, the most prominent peak in the HFD corresponds either to \forb\ (i.e. \dr) or to a multiple of it, which makes the detection recognisable. In addition, some other coincidences occur: for \stari, \Dnu$\sim$11\dr; for \starb\ \Dnu$\simeq$7\dr, and for \starh\ \Dnu$\simeq$5\dr\ (see Table~\ref{tab:sample}). These coincidences might also favour the detection of \dr\ and blur the peak of \Dnu\ in the DFT diagram. Both the HFD of \stari\ (Fig.~\ref{fig:stari}) and the AC show a unique behaviour, displaying several multiples of \dr, maybe caused by all these coincidences. The DFT diagram does not reproduce this in such a clear way. \par

\subsection{Evolved stars}
\label{subsec:evolved}
These stars are \stard\ and \starf. The ratio $(R/\mathrm{R}_{\odot})/(M/\mathrm{M}_{\odot})\simeq2$ for both stars is compatible with that of an evolved star. This means that both of them might show effects of avoided crossings \citep{aizenman1977}, reflected in the oscillation spectrum as a displacement in the frequencies. This shift can break the periodicity, and therefore complicate the analysis.\par

\begin{figure}
	\includegraphics[width=\columnwidth]{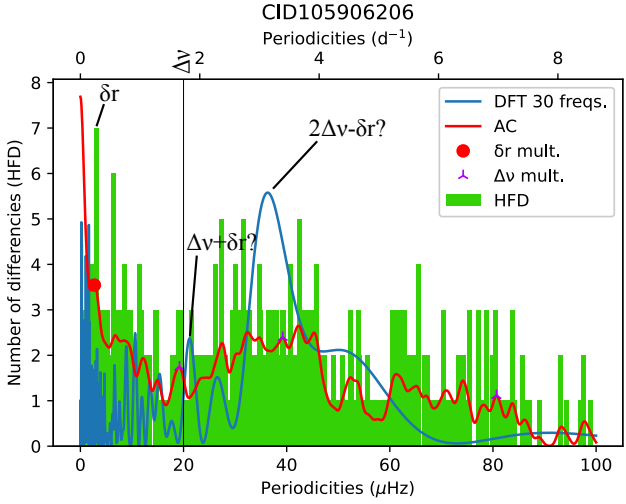}
    \caption{\starf. The peak \Dnu+\dr\ suggested by \citet{paparo2016-I} can be found in the DFT. The avoided crossing might be affecting the position of the peaks and the appearance of the combination 2\Dnu-\dr. \dr\ can be found in both the AC and the HFD}
    \label{fig:starf}
\end{figure}

For \starf, \dr\ can be eye-guessed combined with \Dnu\ in the DFT (Figure \ref{fig:starf}) as \Dnu+\dr. The highest peak could also be 2\Dnu-\dr. Indeed, \citet{paparo2016-I,paparo2016-II} pointed to \Dnu+\dr\ as a possible pattern combination, which would agree with one of the peaks present in the DFT. However, there is no hint of the 2\Dnu-\dr, probably because in those works evolved stars were explicitly dropped from their sample. However, \dr\ can be found in both the HFD and the AC. \par

\begin{figure}
	\includegraphics[width=\columnwidth]{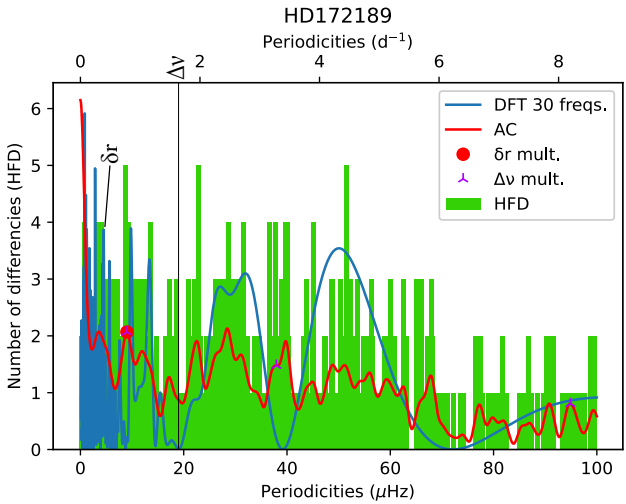}
    \caption{\stard. The peak marked in the AC is the one corresponding to 2\dr\ (and \Dnu/2).}
    \label{fig:stard}
\end{figure}

On the other hand, for the other star of this group, \stard\ we found no pattern combinations (Figure \ref{fig:stard}). We could spot \dr\ as a peak in the DFT, and 2\dr\ (which also is \Dnu/2) was identified in both the HFD and the AC.\par

In any case, not only have we found \dr\ but also it would seem that the combination $A$\Dnu$\pm B$\dr\ is more frequent in evolved stars. In these cases it is difficult to ascertain if the peaks of the Fourier transform and/or the autocorrelation are due to \dr\ alone or to a linear combination with \Dnu.\par

\subsection{Peculiar cases}
\label{subsec:peculiar}
We classified as \emph{peculiar case} the remaining objects of our sample that simply could not be classified in any of the previous groups because their analysis is rather difficult or the results are simply unreliable. These are \starc, \starg\ and \stark.\par

\begin{figure}
	\includegraphics[width=\columnwidth]{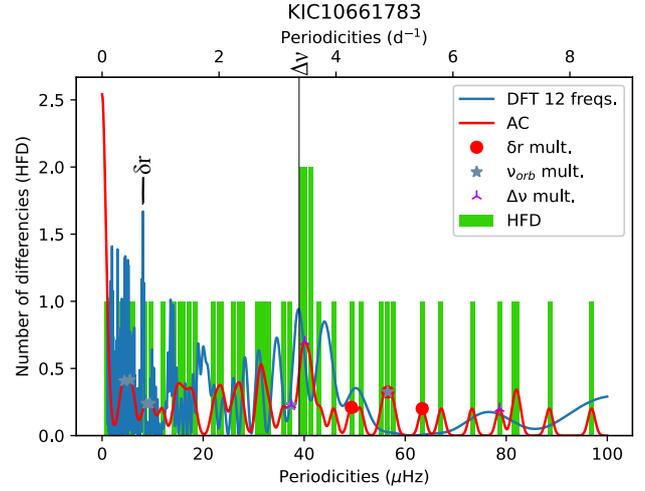}
    \caption{\starc. Despite the low number of frequencies we were able to identify \dr\ as the highest peak in the DFT but not in the HFD nor in the AC. The first multiple of \dr\ that we can find in this last curve is 7\dr. The first peaks in the AC correspond to \forb/2\ and \forb, respectively.}
    \label{fig:starc}
\end{figure}

One of the reasons for a star to be classified as a peculiar case might be the low number of detected oscillation frequencies. This is the case of \starc, with only 12 modes (in the p mode regime). This makes it nearly impossible to make any reliable analysis with our methodology. Even so, we surprisingly found the predicted \dr\ with the DFT diagnostic (Fig. \ref{fig:starc}). This makes DFT the most robust technique for such cases, at least to provide an educated guess on the searched pattern. However, for objects with no previous knowledge, it is difficult to confidently spot the \dr\ peak. For now, we have no explanation on this low number of frequencies, however with a large sample of stars, a similar study could give us more information about the relation between \dr\ and a low number of frequencies, in particular if such sample contains stars with a wide distributions of \dr\ and $i$. Interestingly, \Dnu\ is clearly identifiable in the 3 transforms.\par

\begin{figure}
	\includegraphics[width=\columnwidth]{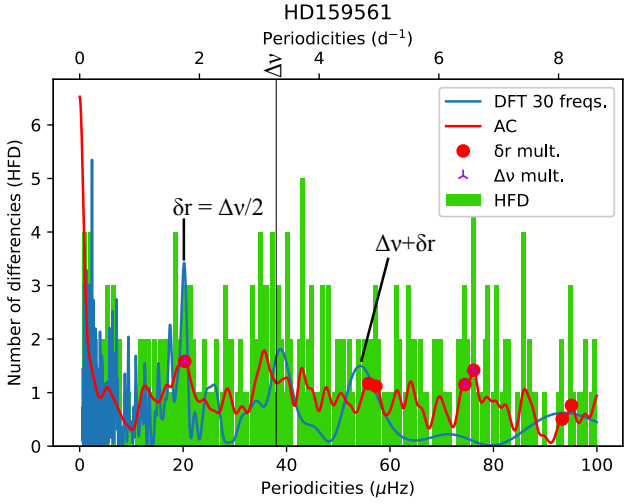}
    \caption{\starg, in which \dr=\Dnu/2. This causes the enhancement of the peak corresponding to that value.}
    \label{fig:starg}
\end{figure}

Another interesting factor that may contribute to modify the distribution of patterns in the oscillation spectrum is fast rotation. With a rotation rate of $\Omega/\Omega_\mathrm{k,\,c}\sim0.6$\footnote{where $\Omega_\mathrm{k,\,c}=\sqrt{{GM}/{R_\mathrm{eq}^3}}$, with $R_\mathrm{eq}$ the stellar equatorial of the current model, as described by \citet{reese2017}}, \starg\ can be considered as a fast rotator. According to \citep{reese2017}, when $\Omega/\Omega_\mathrm{k,\,c}\sim0.7$, which is close enough to the rotation rate of \starg, there is a strong frequency spacing occurrence at \Dnu/2$\sim$\dr, making both of them indistinguishable. This agrees well with the DFT (Figure \ref{fig:starg}). This peak is quite prominent when using the 30 frequencies of highest amplitude and, also, when using 40 frequencies (not shown in the figure). On the other hand, it is interesting to note that the peak corresponding to \Dnu\ only shows when using 30 frequencies. There is also a peak corresponding to the sum of \Dnu\ and \dr.\par

\begin{figure}
	\includegraphics[width=\columnwidth]{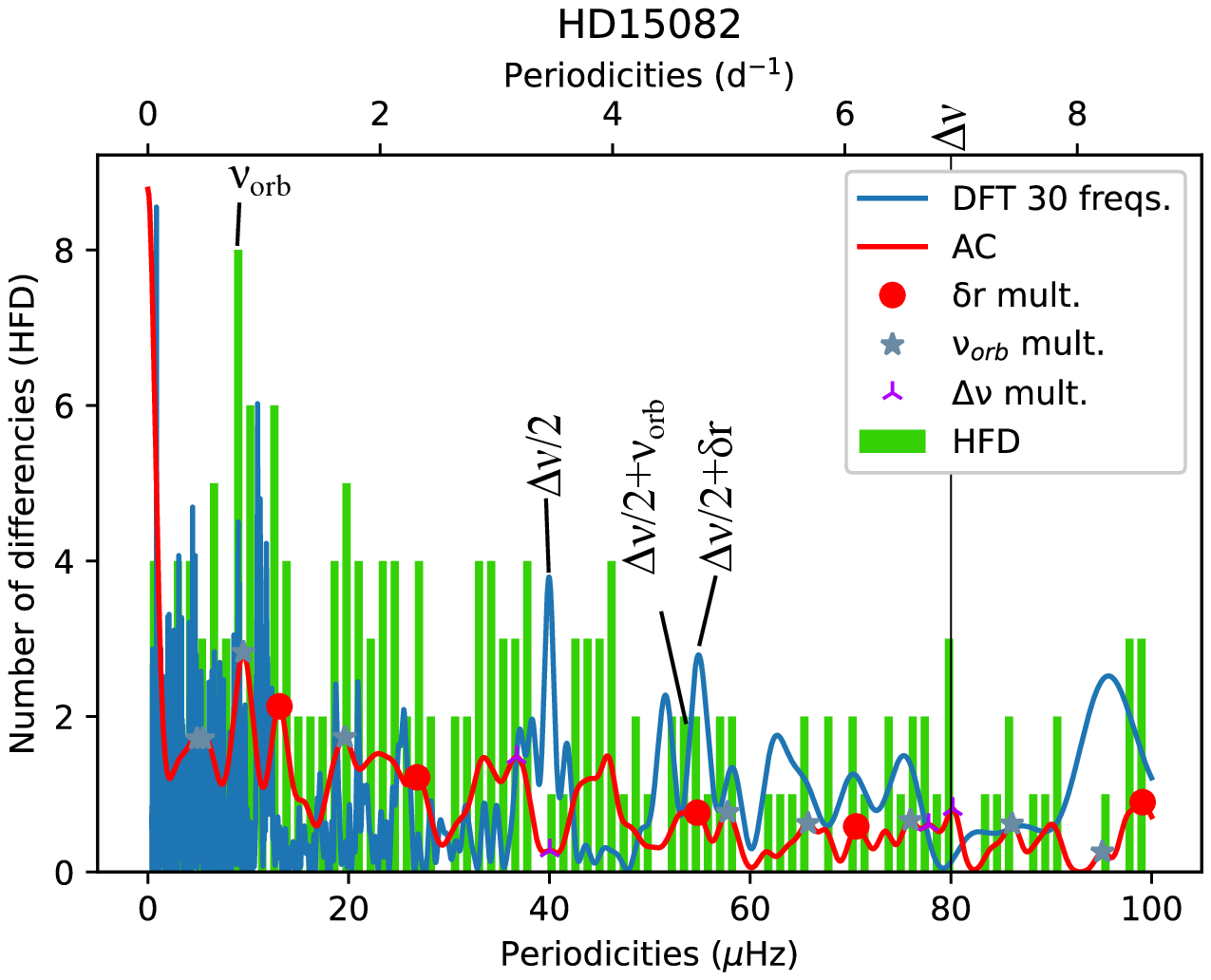}
    \caption{The DFT for \stark\ shows unexpected combinations of \Dnu\ with \forb\ and \dr.}
    \label{fig:stark}
\end{figure}

\stark\ is the only \ds\ star known to have an orbiting planet. In this case the results are quite unexpected because we find a combination of \Dnu/2 and \forb\ plus another one of \Dnu/2 and \dr, along with lots of unidentifiable peaks (see Figure \ref{fig:stark}). In this case, the uncertainty in \Dnu, \dr\ and \vsini\ is one of the highest of all the sample. We think that perhaps the angle of inclination has not been determined accurately enough, and that could be the cause of these unexpected results. This might be due to a misalignment of the orbital and spin axes, as discussed in Sect.~\ref{sec:sample}.\par

\section{Guessing the surface rotation}
\label{sec:unknownrotation}
At last, we have used the results of the previous stars to make an educated guess of \dr\ for three targets of unknown rotation: \stare, \starl\ and \starm.\par

\begin{figure}
	\includegraphics[width=\columnwidth]{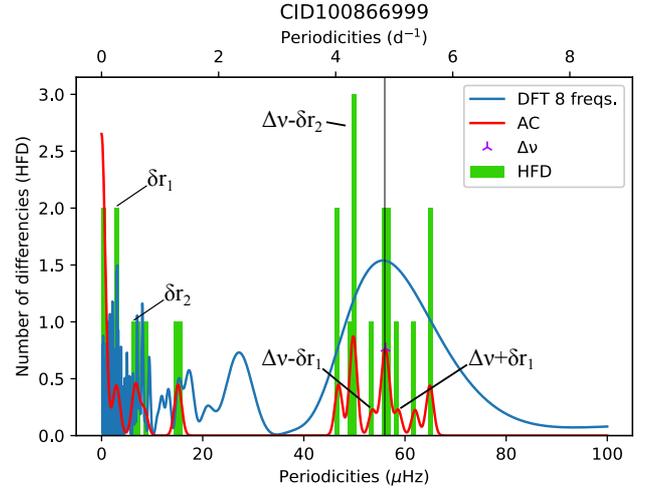}
    \caption{\stare. We have estimated \dr$\sim$6~\muhz\ based on what we have learned from previous cases.}
    \label{fig:stare}
\end{figure}

The star \stare\ has only eight frequencies and this so low number may hamper the analysis (see e.g. the case of \starc\ in Sec.~\ref{subsec:peculiar}). Despite this, we have been able to spot two probable values for \dr: \dr$_1\sim$2.5 \muhz\ and \dr$_2\sim$6.8 \muhz. The highest peak of the HFD (see Fig. \ref{fig:stare}) might correspond to the combination \Dnu-\dr$_2$\ and hence, \dr$_2\sim$6.8 \muhz\ would be consistent with the second maximum of the AC. The other possibility, \dr$_1\sim$2.5 \muhz, corresponds with the first maximum peak of the DFT in the lower frequency region and coincides also with a significant peak in the HFD. This value would explain the symmetric sidelobes around \Dnu, clearly shown by the AC.\par

\begin{figure}
	\includegraphics[width=\columnwidth]{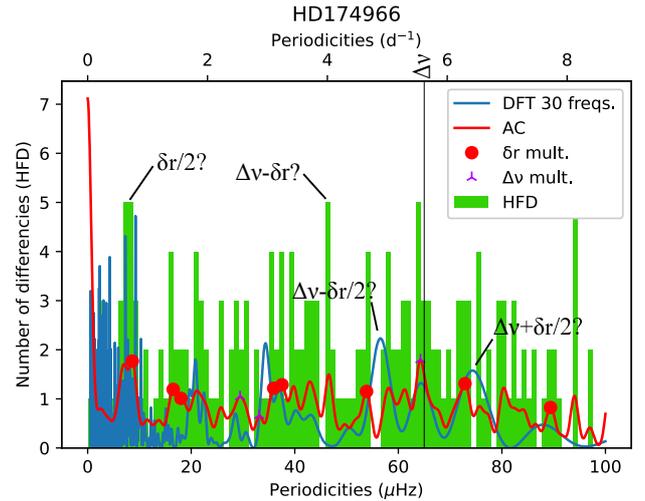}
    \caption{\starl. In the DFT, the peak corresponding to \Dnu/2 might be enhanced because 2\dr$\sim$\Dnu/2. The two peaks surrounding \Dnu\ might as well be a combination of \Dnu$\pm$\dr/2 compatible with \dr$\sim$18-19~\muhz. In the HFD, the first peak around 8-9~\muhz\ could well be \dr/2. There is also another peak that could correspond to \Dnu-\dr.}
    \label{fig:starl}
\end{figure}

For \starl, \citet{GH13} have estimated \Dnu=65 \muhz\ and \dr$\sim$18 \muhz, guessing the stellar inclination from line-profile variations. Also, from the data in Table \ref{tab:sample}, extracted from that work, we can derive a range for $\Omega/\Omega_\mathrm{k,\,c}$ between 0.31 and 0.42. \citet{reese2017} found that 2\dr$\sim$\Dnu/2 when $\Omega/\Omega_\mathrm{k,\,c}\sim0.3$ (thus enhancing these peaks), which could be the case here. In the DFT (Figure \ref{fig:starl}) there is a peak around 34 \muhz\ that would well agree with all these assumptions. This DFT also shows two peaks around \Dnu\ which could be some sort of combination of the type \Dnu$\pm$\dr/2. This would be compatible as well with a peak in the HFD guessed as \Dnu-\dr. That would give an estimated value of \dr$\sim$19 \muhz, quite in accordance with the aforementioned 18 \muhz.\par

\begin{figure}
	\includegraphics[width=\columnwidth]{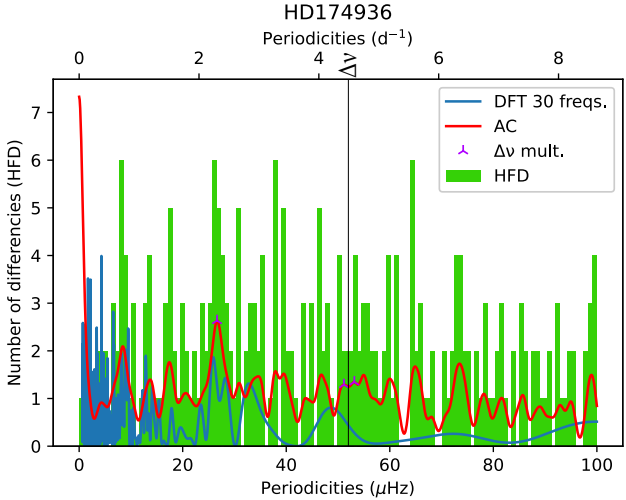}
    \caption{\starm. The high peak at \Dnu/2 in the DFT, as well as a peak at about half this value, suggests this star might be a fast rotator of the type \Dnu/2$\sim$\dr. In the HFD, the first peak at about 8 \muhz\ is a bit low to concur with the finding in the DFT. But there is another one, corresponding to \Dnu/2, which is equally prominent and would be coherent with our assumption.}
    \label{fig:starm}
\end{figure}

In \starm\ we have no information at all about the inclination angle. We can only establish that $\Omega/\Omega_\mathrm{k,\,c}\geq0.42$ and \dr\ $\geq$ 17 \muhz (Table~\ref{tab:sample}). From \citet{GH09} we have \Dnu=52 \muhz. If this star were a fast rotator in which $\Omega/\Omega_\mathrm{k,\,c}\sim0.7$, like \starg\ discussed in section \ref{subsec:peculiar}, we would expect that \Dnu/2 $\simeq$\dr\ and that peak would stand out in the DFT. As we can see in the Figure \ref{fig:starm}, there is a peak in the DFT, and another one in the HFD which would well confirm this assumption.\par

\section{Conclusions}
\label{sec:conc}

We have used the discrete Fourier transform, the autocorrelation function and the histogram of frequency differences in order to search for a pattern corresponding to the rotational splitting (\dr). We used a selected sample of \dss\ in eclipsing binary systems taken from \citet{GH17}. The procedure consisted in searching for multiples of \dr\ and linear combinations with the low-order \Dnu\ in the calculated functions. Our main conclusion is that the primary identification of \Dnu\ is crucial to achieve a positive result.\par

We have been able to find \dr\ for the majority of the studied stars, mainly from the combined analysis of the Fourier transform and the autocorrelation function, as predicted by \citet{reese2017}. We have found that the histogram of frequency differences mimics the behaviour of the autocorrelation, resulting in redundant information. Hence, we can claim that the signature of \dr\ is present and identifiable using asteroseismological data only. In fact, we were able to predict the splitting for three stars for which no clue on the rotational velocity was given in the literature.\par

The results using the discrete Fourier Transform or the autocorrelation function are not very different (although they complement each other), as expected from the \mbox{Wiener-Khinchin} theorem, which states that ``The autocorrelation function of a wide-sense-stationary random process has a spectral decomposition given by the power spectrum of that process.''\par

For the sample of stars studied here, we have classified the stars in different subsets regarding the behaviour of the diagnostic functions. The characteristics of these subsets seem to have to do with the evolutionary stage of the stars, their rotation and/or the number of extracted frequencies. For example, the group of evolved stars seems to privilege a linear combination of \Dnu\ and \dr, whereas in the other cases we can identify directly a multiple of \dr.\par

In any case, this conclusion may be considered with caution since the number of objects in the sample is small. A larger sample would allow to develop an automated methodology to spot the rotational splitting in the oscillation spectrum of any \dss.\par

We tried analysing the 30 frequencies with the lowest amplitude but the results were inconclusive. It is important to note that the method used to obtain the frequencies is crucial to perform the analysis. The classical pre-whitening technique used in asteroseismology to extract the frequencies from the light curve may introduce spurious frequencies \citep[][]{baran2015} of low amplitude. Recently, \citet{defranciscis2019} proved that this issue can be solved considering the hypotesis that the light curve consists of a superposition of harmonic modes together with fractal background noise. These authors developed a self-consistent technique to stop the pre-whitening before these spurious frequencies are introduced. In a future work, this method could be used to extract the frequencies of the sample of stars used in this study, thus allowing the analysis of the lowest amplitude frequencies.

In the near future, we plan to study these results from a theoretical point of view, using perturbative and non-perturbative oscillations. This would allow us to analyse the influence of rotation and evolutionary stage on the diagnostic diagrams studied here and therefore to improve the methodology to find the rotational splittings. In a more distant future, a good determination of \Dnu\ and \dr\ will be key in identifying the oscillation modes and therefore to better study their visibilities and physical amplitudes.\par

\section*{Acknowledgements}

The authors would like to thank the referee Daniel Reese for his insightful comments.\par

ARB, JPG and RG acknowledge funding support from Spanish public funds for research from project ESP2017-87676-C5-5-R from the `Plan Estatal de Investigaci\'on Cient\'ifica y T\'ecnica y de Innovaci\'on', from project PID2019-107061GB-C63 from the `Programas Estatales de Generaci\'on de Conocimiento y Fortalecimiento Cient\'ifico y Tecnol\'ogico del Sistema de I+D+i y de I+D+i Orientada a los Retos de la Sociedad', and from the State Agency for Research through the ``Center of Excellence Severo Ochoa'' award to the Instituto de Astrof\'isica de Andaluc\'ia (SEV-2017-0709), all from the Spanish Ministry of Science, Innovation and Universities (MCIU).\par

ARB also acknowledges funding support from project PRE2018-084322 from the `Programa Estatal de Promoci\'on del Talento y su Empleabilidad del Plan Estatal de Investigaci\'on Cient\'ifica y T\'ecnica y de Innovaci\'on 2013-2016' of the Spanish MCIU.\par

AGH acknowledges funding support from Spanish public funds for research under projects ESP2017-87676-2-2 the Spanish Ministry of Science and Education, and from ``FEDER/Junta de Andaluc\'{\i}a-Consejer\'{\i}a de Econom\'{\i}a y Conocimiento'' under project E-FQM-041-UGR18 by Universidad de Granada.\par

JCS acknowledges funding support from Spanish public funds for research under projects PID2019-107061GB-C63 and ESP2017-87676-2-2, and from project RYC-2012-09913 under the `Ram\'on y Cajal' program of the Spanish Ministry of Science and Education.\par

\section{Data availability}
The data analysed for each star in this article are available in the respective articles cited in \citet{GH17}.\par




\bibliographystyle{mnras}
\bibliography{library} 




\appendix
\section{Remaining figures}
\label{sec:appendix}

\begin{figure}
	\includegraphics[width=\columnwidth]{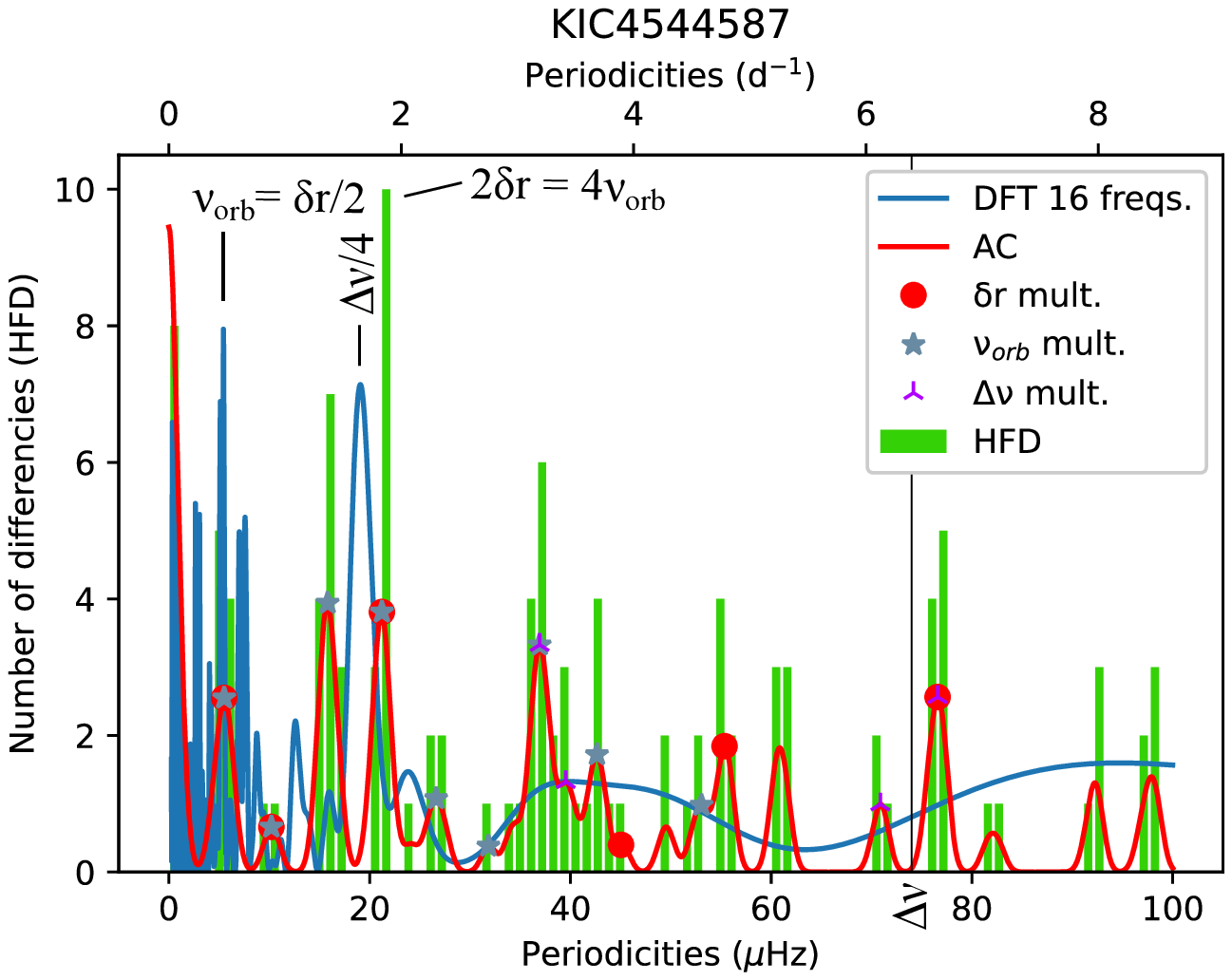}
    \caption{\starb. In the DFT, the peak corresponding to \forb\ might be enhanced because \dr$\sim$2\forb. There is also a peak corresponding to \Dnu/4. In the HFD, the peak around 22~\muhz\ (2\dr\ or 4\forb) also stands out.}
    \label{fig:starb}
\end{figure}

\begin{figure}
	\includegraphics[width=\columnwidth]{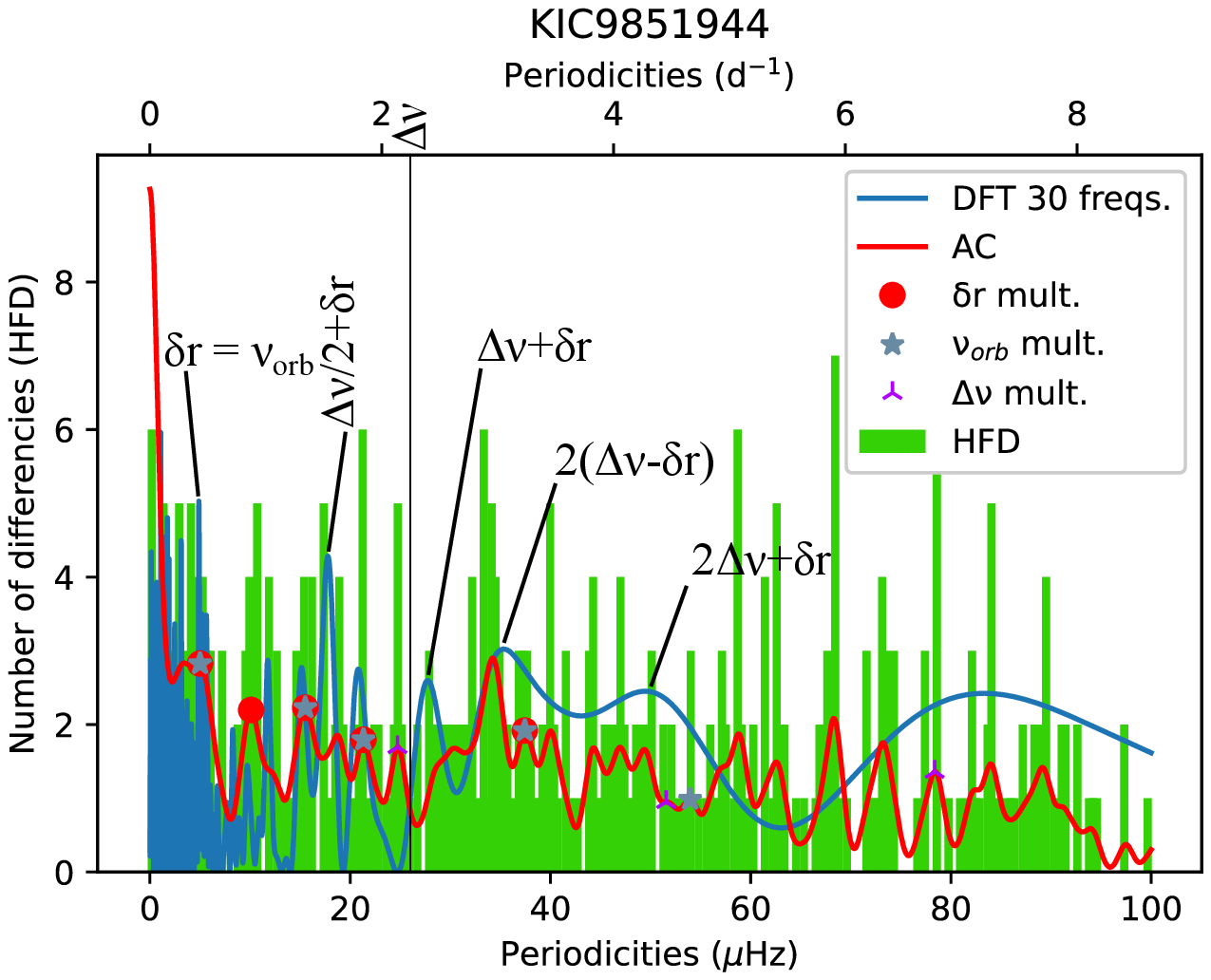}
    \caption{\starh. In this star \dr=\forb. Several combinations of \Dnu\ and \dr\ (or \forb) stand out in the DFT. The AC and the HFD also show some multiples of \dr\ and \forb.}
    \label{fig:starh}
\end{figure}

For the sake of completeness, we include here the two remaining figures that have not been cited in the main text: \starb\ (Fig. \ref{fig:starb}) and \starh\ (Fig. \ref{fig:starh}).

\bsp	
\label{lastpage}
\end{document}

%% file: personal_set.tex
 \usepackage{multirow}

\newcommand{\logg}{$\log g$}

\newcommand{\ds}{$\delta$~Sct}
\newcommand{\dss}{$\delta$~Sct stars}

\newcommand{\corot}{{\it CoRoT}}

\newcommand{\kepler}{{\it{Kepler}}}
\newcommand{\tess}{{\it{TESS}}}
\newcommand{\most}{{\it{MOST}}}

\newcommand{\vsini}{$v\sin\,i$}
\newcommand{\kms}{$\mathrm{km}~\mathrm{s}^{-1}$}
\newcommand{\teff}{T$_{\mathrm{eff}}$}
\newcommand{\msol}{$\mathrm{M}_{\odot}$}
\newcommand{\rsol}{$\mathrm{R}_{\odot}$}

\newcommand{\muhz}{$\mu\mbox{Hz}$}

\newcommand{\Dnu}{$\Delta\nu$}
\newcommand{\dr}{$\delta r$}

\newcommand{\ele}{$\ell$}

\newcommand{\gh}{Garc\'{\i}a Hern\'andez}

%% file: table_01.tex
    \resizebox{\textwidth}{!}{\begin{tabular}{lccccccccccc}
        \hline
        Star         & \teff     & \logg   & M         & R         & \Dnu    & \vsini & i              & \dr     & \forb        & \Dnu/\dr   & N. of \\
                     &  (K)      &         & (\msol)   & (\rsol)   & (\muhz) & (\kms) & (\textdegree)  & (\muhz) & (\muhz)      &            & freqs \\
        \hline                                                                                     
        \stara$^a$   & 6606(70)  & 3.74(1) & 1.86(4)   & 3.05(1)   & 29(1)   & 25.7(1.5) & 88.176(2)   & 1.9(1)  & 0.445980(1)  & 15.3       & 400 \\
        \starj$^a$   & 7480(200) & 4.1(1)  & 1.9(1)    & 2.1(2)    & 52(1)   & 14.4(1.4) & 68(3)       & 1.7(2)  & 0.75468(1)   & 30.6       & 321 \\
        \starb$^b$   & 7750(180) & 4.33(1) & 1.61(6)   & 1.57(3)   & 74(1)   & 76(15)    & 88(3)       & 11(2)   & 5.28715(1)   & 6.7        & 16 \\
        \starh$^b$   & 6902(100) & 3.69(3) & 1.79(7)   & 3.16(4)   & 26(1)   & 72.1(9)   & 74.52(2)    & 5.3(1)  & 5.348706(2)  & 4.9        & 52 \\
        \stari$^b$   & 9128(130) & 4.28(2) & 1.96(6)   & 1.67(4)   & 77(1)   & 50.7(9)   & 75.178(2)   & 7.2(2)  & 7.17542974(4) & 10.7      & 60 \\
        \stard$^c$   & 7750(-)   & 3.48(8) & 1.8(2)    & 4.0(1)    & 19(1)   & 78(3)     & 73.2(6)     & 4.6(2)  & 2.02983(1)   & 4.1        & 50 \\
        \starf$^c$   & 6750(150) & 3.53(1) & 2.24(4)   & 4.22(2)   & 20(2)   & 47.8(5)   & 81.42(13)   & 2.61(3) & 3.13272(1)   & 7.7        & 202 \\
        \starc$^d$   & 7764(54)  & 3.938(4) & 2.10(3)  & 2.56(2)   & 39(1)   & 78(3)     & 82.4(2)     & 7.0(3)  & 9.399407(2)  & 5.6        & 12 \\
        \starg$^d$   & 8047(154) & 3.9(2)  & 2.4(4)    & 2.69(1)   & 38(1)   & 239(12)   & 87.5(6)     & 19(1)   & 0.0036762(1) & 2.0        & 40 \\
        \stark$^d$   & 7430(100) & 4.3(2)  & 1.49(3)   & 1.43(4)   & 80(2)   & 90(10)    & 87.7(1.8)   & 14(2)   & 9.48798(1)   & 5.7        & 64 \\
        \stare$^e$   & 7300(250) & 4.1(1)  & 1.8(2)    & 1.9(2)    & 56(1)   & ---       & 80(2)       & ---     & 4.12069(3)   & ---        & 8 \\
        \starl$^e$   & 7555(50)  & 4.22(1) & 1.515(15) & 1.58(3)   & 65(1)   & 126.1(1.2) & 62.5(+7.5-17.5)     & 18(-)   & ---  & 3.6       & 172 \\
        \starm$^e$   & 8000(200) & 4.1(2)  & 1.65(26)  & 1.97(32)  & 52(-)   & 169.7(-)  & ---         & ---     & ---          & ---        & 422 \\
        \hline
    \end{tabular}}
    \caption{Main physical parameters of the stars analysed in this work. The error is enclosed between brackets and the unit refers to the last digit of the corresponding value except when a decimal point is used. The data have been taken from their respective paper. The last column shows the number of frequencies extracted for each star. The stars are grouped as in section \ref{sec:results}. a: Favourable cases (Sec. \ref{subsec:favourable}). b: Orbital Synchronisation (Sec. \ref{subsec:synchro}). c: Evolved stars (Sec. \ref{subsec:evolved}). d: Peculiar cases (Sec. \ref{subsec:peculiar}). e: Unknown rotation (Sec. \ref{sec:unknownrotation}). }
    \label{tab:sample}
	